\documentstyle[iopconf1]{article}
\begin{document}
\begin{flushright}
UMD-PP-98-04
\end{flushright}

\title{Left-Right Symmetry Just Beyond MSSM, 
Electric Dipole Moment of the Neutron and HERA Leptoquarks}

\author{R. N. Mohapatra\dag\footnote{E-mail:
rmohapatra@umdhep.umd.edu}}

\affil{\dag\ Department of Physics, University of Maryland, College Park,
MD-20742}

\beginabstract
The supersymmetric left-right (SUSYLR) models solve many of the
problems of the minimal supersymmetric standard model (MSSM) such as the
R-parity and SUSYCP problems. The first one implies that supersymmetry can
provide a naturally
stable LSP (lightest supersymmetric particle) which can serve as the cold
dark matter of the universe. It is then shown that if
$W_R$ mass in the TeV range, the SUSYLR models also 
provide a natural solution to the strong
CP problem without the need for an axion. It is therefore argued that
the theory just beyond the MSSM is the SUSYLR model. A crucial prediction of
the model is the prediction of the electric dipole moment of the
neutron of $d^e_n\simeq 10^{-25}-10^{-26}$ ecm arising from one-loop
contribution to the strong CP parameter $\bar{\theta}$. Other predictions
are a light doubly charged scalar boson and its fermionic superpartner
with masses in the few hundred GeV range. Finally,
it is pointed out how a simple extension of the model incorporates
leptoquarks that can explain the HERA anomaly without giving up R-parity
violation
\endabstract

\section{Introduction}

One of the fundamental new symmetries of nature that has been the
subject of intense discussion in particle physics of the past decade
is the symmetry between bosons and fermions, known as
supersymmetry.  In addition to
the obvious fact that it provides the hope of an unified
understanding of the two known forms of matter, the bosons and
fermions, it has also provided a mechanism to solve two conceptual
problems of the standard model, viz. the possible origin of the
weak scale as well as its stability under quantum corrections. The
recent developments in strings, which embody supersymmetry in an
essential way also have the potential to lead to 
an ultimate theory of everything. It is
therefore naturally a common belief among particle physicists that
the next step beyond the standard model is the supersymmetric standard
model where the supersymmetry of the model is
softly broken by some hidden sector mechanism. The 
nature and origin of supersymmetry breaking is not relevant for the
discussion of what follows. We will therefore make only minimal
assumptions about it in this article.

We will first review the basic features of the minimal supersymmetric
extension of the standard model (MSSM) and argue that while two of the most
critical problems of the standard model i.e. the stability of the Higgs mass
and the generation of electroweak symmetry breaking are solved in the
MSSM, it creates several new problems which standard model solved in an
elegant manner. The new problems include arbitrary amount of baryon and lepton
number violation, large electric dipole moment of the neutron among others.
The former unfortunately means that MSSM does not provide a natural candidate
for cold dark matter of the universe, that is generally considered another
 major selling point for supersymmetry. 
We then argue that if the MSSM is assumed to arise as the low energy limit
of a supersymmetric left-right model (SUSYLR), then R-parity arises as
a natural symmetry keeping baryon and lepton number as automatic
symmetries as in the standard model and furthermore it also provides
a natural solution to the large electric dipole moment problem of the
MSSM (also known as the SUSYCP problem).
The lightest supersymmetric particle (LSP) is
then a naturally stable particle and can act as the CDM of the universe.
More interestingly, we find that if the scale of the righthanded symmetry
is in the TeV range, the model also provides a solution to the strong CP
problem without the need for an axion, thereby solving a problem of both
the standard model as well as the MSSM. A crucial prediction of the theory
is that the dipole moment of the neutron arising from the one loop finite
contribution to the $\bar{\theta}$ parameter is of order $10^{-25}-10^{-26}$
ecm. This is accesible to the next generation of experments searching for the
edm of neutron. This idea therefore has a very good chance to tested in the
next decade. 

\section{The MSSM and its problems}
 
The MSSM is the minimal supersymmetric extension of the standard 
model\cite{nath}
and is based on the same gauge group as the standard model i.e. 
$SU(3)_c\times SU(2)_L\times U(1)_Y$.
 In Table~\ref{ss!1}, we give the particle content of the
model.

\begin{table}
\caption{The particle content of the supersymmetric standard
model. For matter and Higgs fields, we have shown the left-chiral
fields only. The right-chiral fields will have a conjugate
representation under the gauge group.\label{ss!1}}
\begin{center}
\begin{tabular}{|c||c||c||c|}
\hline\hline 
 Superfield & Particles & Superpartners & gauge \\ 
& & &  transformation \\ \hline\hline
 Quarks $Q$ &   $(u, d)$  & $(\widetilde{u}, \widetilde{d})$ & $(3,
2, {1\over 3})$\\  
 Antiquarks $ U^c$ & $u^c$ & $\widetilde{ u^c}$ & $(3^*,
1, - {4\over 3})$ \\ 
Antiquarks  $D^c$ & $d^c$ & $\widetilde{d^c}$ & 
$(3^*, 1, {2\over3})$ \\ 
 Leptons $L$ & $(\nu, e)$ & $(\widetilde{\nu}, \widetilde{e})$ 
& $(1, 2 -1)$ \\ 
 Antileptons  $E^c$ & $e^c$  & $\widetilde{e^c}$ & $(1,
1, 2)$ \\  
 Higgs Boson $\bf H_u$ & $(H^+_u, H^0_u)$ & $(\widetilde{H}^+_u,
\widetilde{H}^0_u)$ & $(1, 2, +1)$ \\ 
Higgs Boson $\bf H_d$ & $(H^0_d, H^-_d)$ & $(\widetilde{H}^0_d,
\widetilde{H}^-_d)$ & $(1, 2, -1)$ \\ 
Color Gauge Fields & $G_a$ & $\widetilde{G}_a$ & 
$(8, 1, 0)$ \\ 
Weak Gauge Fields & $W^{\pm}$, $Z$ & $\widetilde W^{\pm},
\widetilde{Z}$ & \\  
Photon & $\gamma$ & $\widetilde{\gamma}$ &   \\ \hline\hline
\end{tabular}

\end{center}
\end{table}
The first point to note is that while the gauge interaction
of the standard model fermions remains unchanged in this supersymmetrized 
version,
 the weak interactions of the squarks and the sleptons are very different from
their fermionic partners due to supersymmetry
breaking. This has the phenomenological implication that the
gaugino-fermion-sfermion interaction changes generation leading to
potentially large flavor changing neutral current (FCNC) effects such as
$K^0$-$\bar{K}^0$ mixing, $\mu\to e\gamma$ decay etc unless the
sfermion masses of different generations are chosen to be very close
in mass. This is the FCNC problem of the MSSM and we do not dwell
on this here since the SUSYLR model does not throw light on this.

To discuss the other problems of MSSM,
let us discuss the superpotential of the
model.  It consists of two parts:
	\begin{eqnarray}
W= W_1 + W_2 \,,
	\end{eqnarray}
where
	\begin{eqnarray}
W_1 &=& h^{ij}_{\ell}E^c_i L_j {\bf H_d} + h^{ij}_d Q_i 
D^c_j {\bf H_d} + 
h^{ij}_u Q_i U^c_j {\bf H_u} + \mu {\bf H_u H_d} \\ 
W_2 &=& \lambda_{ijk} L_iL_jE^c_k +\lambda'_{ijk} L_iQ_jD^c_k 
+\lambda''_{ijk}  U^c_i D^c_j D^c_k \,,
\label{ss.W}
	\end{eqnarray}
$i,j,k$ being generation indices. 
Note that while the terms in $W_1$ conserve baryon and lepton
number those in $W_2$ do not and yet they are perfectly
allowed by the gauge invariance of the model. Thus baryon and
lepton number are no more natural symmetries of the model as they were
for the standard model. The latter are known as the
$R$-parity breaking terms where $R$-parity is defined as
$R = (-1)^{3(B-L)+2S}$ where $S$ is the spin of the particle.
These R-parity violating couplings are severely constrained by
present experiments\cite{gautam}. For instance,
present experimental limits on the lifetime of the proton imply that
the product $\lambda^{\prime}\lambda''\leq 10^{-24}$ which is much too severe a
fine tuning. Furthermore the LSP in this model is too shortlived to play the
role of CDM of the universe.
 
In order to prevent the appearance
of these terms, one generally assumes the existence of some global symmetry
(which could be discrete or continuous). First of all this is an additional
assumption which diminishes the appeal of MSSM somewhat; but perhaps more
serious is the lore that nonperturbative quantum gravity tends to break all
global symmetries of nature making the appearance of Planck suppressed
$\Delta B\neq 0$ and $\Delta L\neq 0$ in the low energy Lagrangian quite
plausible\cite{gid88}. Even though
in general the strengths of these terms are small due to the Planck scale
suppression, they are not small enough to let the LSP live longer than 
the age of the universe. Thus it may be preferable to search for more
natural ways to eliminate the R-parity violating terms from the supersymmetric
models. It was realized in
mid-80's \cite{moha86} that such is the case in the supersymmetric
version of the left-right model that implements the see-saw
mechanism. We demostrate this point in the next section.

A second problem with the MSSM lies in its predictions for the CP
violating effects being too large. To see this note that in the simple
versions of MSSM, there are at least three phases including the 
usual KM-phase $\delta_{KM}$ of the standard model. The two of the
extra phases reside in the soft breaking parameter $A$ and the Higgs
mixing mass $\mu$. A priori, all phases in a theory are arbitrary. In the
standard model even though the KM phase can be left arbitrary, its physical
effects are always suppressed by the product of various quark mixing angles
leading to natural suppression of all CP violating effects. In the MSSM 
however, such suppressions do not arise in all CP violating effects. For
instance, if we calculate the electric dipole moment of the neutron, then
we typically get an expression of the form:
\begin{eqnarray}
d^e_n\simeq \frac{e}{16\pi^2}\frac{m_d}{M^4_{\tilde{q}}} Arg(m_{\tilde{g}}
[A-\mu tan\beta])
\end{eqnarray}
A simple evaluation of the above down quark electric dipole moment leads to the
conclusion that unless either (i) the squark masses are of order 3 TeV or
(ii) $Arg({m_{\tilde{g}}A})$ and $Arg(m_{\tilde{g}}\mu)$ are less than $10^{-3}$
if squark masses $M_{\tilde{q}}\simeq 100 $ GeV, the edm of the neutron will
come out to be three orders of magnitude higher\cite{garisto} 
than the present experimental
upper bound. In either case, we have a fine tuning problem for the
theory, the very problem supersymmetry was supposed to solve. In the first
case one has to fine tune to get the Higgs mass of order $m_W$ and in the
second case, the new phases of the model (unlike the CP phase of the standard
model) has to be tuned down by three orders of magnitude from its natural
value. 

The third and final problem of MSSM that we discuss here 
is the strong CP problem. It is by now quite well-known
that the Quantum Chromodynamics theory of strong interactions
which has been so successful in its description of low and high
energy hadronic phenomena implies an arbitrary amount of CP-violation
in the $\Delta S= 0$ processes. This results from the 
periodic vacuum structure of QCD and is given by the following effective
Lagrangian\cite{jkim87}:
\begin{eqnarray}
L_{\theta}={\theta}\frac{g^2_s}{32\pi^2} G_{\mu\nu}\tilde{G}^{\mu\nu}
\end{eqnarray}
This has the disastrous implication that it leads to an electric dipole
moment of the neutron given by $d^e_n\simeq 10^{-16} \theta$ ecm. 
The present upper limits on the $d^e_n\leq 10^{-25}$ ecm then imply that
$\theta \leq 10^{-9}$ which requires a severe fine tuning of the parameters
and is therefore not acceptable in a final theory. The inclusion of the
electroweak sector of the theory leads to a slight modification of the above
results: the actual observable $\theta$ is called $\bar{\theta}\equiv
\theta + Arg Det M_{q}$. The above upper bound actually applies to 
$\bar{\theta}$.

What goes wrong if we set $\bar{\theta}=0$ by hand in a given theory ?
The electroweak sector of the theory then generates an infinite contribution
to $\bar{\theta}$ in loop levels requiring that we set the $\bar{\theta}$
to a small value in every order of perturbation theory. For instance,
in the standard model, one finds a nonzero and infinite $\bar{\theta}$
at the sixth loop level\cite{ellis}.
Coming to the MSSM, since one now has colored gluinos, the effective
parameter describing strong CP violation is now given by:
\begin{eqnarray}
\bar{\theta}= \theta + Arg Det M_q -3 Arg M_{\tilde{G}}
\end{eqnarray}
We see that the same gluino phase that was responsible for large edm of
neutron at the one loop level now contributes also to the $\bar{\theta}$
at the tree level. If we set the gluino phase to zero at the tree level
by hand, we will get an infinite value for the phase in the MSSM at the two
loop level\cite{duga85}(see Figure 1). 
Thus one could argue that in the MSSM, the strong CP
problem is worse than in the standard model.

The most popular solution to the strong {\it CP} problem\cite{jkim87} 
is the Peccei-Quinn solution which requires the complete
gauge theory of electroweak and strong interactions to respect a 
global chiral $U(1)$ symmetry. 
This symmetry must however be
spontaneously broken in the process of giving mass to the
W-boson and fermions leading 
to a pseudo-Goldstone boson in the particle spectrum known in
the field as the axion. There are three potential problems with this otherwise
beautiful proposal: (i) the axion has not been experimentally discovered 
as yet and the window is closing in on it; and (ii) if 
non-perturbative gravitational effects induced by black holes and wormholes
are important in particle physics as is believed by some\cite{kami92},
then the axion solution would require fine tuning of the gravitationally
induced couplings by some 50 orders of magnitude. This will make the axion 
theory quite contrived. Finally, it is not easy to get the correct scale
for the axion in superstring models which in the opinion of many people
is the ultimate theory of all matter and forces.

A second class of solutions that does not lead to any near massless boson  
is to require the theory to be invariant under discrete 
symmetries\cite{mbeg78,geor78}. The most physically
motivated of such theories are the ones\cite{mbeg78} based on the left-right
symmetric theories of weak interactions\cite{pati74}.
These theories are based on the gauge group
 $SU(2)_L\times SU(2)_R\times U(1)_{B-L}$ 
with quarks and leptons assigned in a left-right symmetric manner.
Such models are also completely quark-lepton symmetric. To see
how parity symmetry of the Lagrangian helps to solve the strong 
{\it CP} problem, let us note that invariance under parity sets $\theta=0$ 
because $G \tilde{G}$ is odd under parity.
Additionally, constraints of left-right symmetry imply that the Yukawa
couplings of quarks responsible for the generation of quark masses
are hermitean. If furthermore the vacuum expectation values (VEVs) of the
Higgs fields responsible are shown to be real, then this would
automatically lead to $\bar{\theta}=0$ at the tree level. If the one loop
contributions also preserve the hermiticity of the quark mass matrices,
then we have a solution to the strong {\it CP}
problem. In the nonsupersymmetric left-right models with nontrivial
{\it CP} violation, it is well known that in general VEVs of
the Higgs field are not real. This, in the past led to suggestions that
either new discrete symmetries be invoked together with left-right symmetry
or new vectorlike fermions be added to the theory\cite{mbeg78}. 
Such theories also do not suffer from the Planck scale implied 
fine tunings\cite{bere93}. It always
remained a challenge to solve the strong {\it CP} problem using 
only left-right
symmetry since often new additional symmetries 
invoked are not motivated from any
other consideration. Recently it has been shown that left-right symmetry
in combination with supersymmetry solves the strong CP problem without
the axion.

It is the goal of this talk to discuss how the dark matter, SUSY CP
as well as the strong CP problems are all solved without any extra assumptions
and introducing any extra fermions
if the MSSM arises as a low energy limit of the supersymmetric left-right
(SUSYLR) model with the Higgs structure required to obtain the see-saw
mechanism for the neutrino masses. Furthermore solution to the strong
CP problem requires that the scale of $SU(2)_R$ symmetry breaking must
be in the TeV range. A crucial prediction of the model is a dipole moment
of the neutron $~ 10^{-25}-10^{-26}$ ecm, which in the range accessible to
the next generation of the proposed experiments.  

\section{Supersymmetric Left-Right model}

The left-right symmetric theories are based on gauge group  
SU$(2)_L \, \times$ SU$(2)_R \, \times$ U$(1)_{B-L}$ with quarks and 
leptons transforming as doublets under SU$(2)_{L,R}$. These models were
originally proposed\cite{pati74} in order to understand the origin of
parity violation as a consequence of spontaneous symmetry breaking
rather than as an intrinsic part of the gauge interactions. It was then
used to develop models of CP violation as well as to discuss the strong CP
problem\cite{mbeg78}. The version of the model we will be interested
in was proposed by G. Senjanovi\'c and this author \cite{mohagoran80}
in order to implement the see-saw mechanism to understand the small
neutrino masses. This see-saw requirement fixes the Higgs structure
of the model and this is the version which seems to have all the desirable
properties that we are interested in here.
 
In Table 2, we denote the quark, lepton and Higgs 
superfields in the theory along with their transformation properties
under the gauge group. Note that we have chosen two bidoublet fields
to obtain realistic quark masses and mixings (one bidoublet implies 
a Kobayashi-Maskawa matrix proportional to unity, because supersymmetry forbids
$\tilde{\Phi}$ in the superpotential).

\begin{table}
\begin{tabular}{|c|c|} \hline
Fields           & SU$(2)_L \, \times$ SU$(2)_R \, \times$ U$(1)_{B-L}$ \\
                 & representation \\ \hline
Q                & (2,1,$+ {1 \over 3}$) \\
$Q^c$            & (1,2,$- {1 \over 3}$) \\
L                & (2,1,$- 1$) \\
$L^c$            & (1,2,+ 1) \\
$\Phi_{1,2}$     & (2,2,0) \\
$\Delta$         & (3,1,+ 2) \\
$\bar{\Delta}$   & (3,1,$- 2$) \\
$\Delta^c$       & (1,3,+ 2) \\
$\bar{\Delta}^c$ & (1,3,$- 2$) \\ \hline
\end{tabular}
\caption{Field content of the SUSY LR model}
\end{table}

The superpotential for this theory is given by (we have suppressed
the generation index):

\begin{eqnarray}
W & = & 
{\bf Y}^{(i)}_q Q^T \tau_2 \Phi_i \tau_2 Q^c +
{\bf Y}^{(i)}_l L^T \tau_2 \Phi_i \tau_2 L^c 
\nonumber\\
  & +  & i ( {\bf f} L^T \tau_2 \Delta L + {\bf f}_c 
{L^c}^T \tau_2 \Delta^c L^c) 
\nonumber\\
  & +  & \mu_{\Delta} {\rm Tr} ( \Delta \bar{\Delta} ) + 
\mu_{\Delta^c} {\rm Tr} ( \Delta^c \bar{\Delta}^c ) +
\mu_{ij} {\rm Tr} ( \tau_2 \Phi^T_i \tau_2 \Phi_j ) 
\nonumber\\
 & + & W_{\it NR}
\label{eq:superpot}
\end{eqnarray}  
where $W_{\it NR}$ denotes non-renormalizable terms arising from
higher scale physics such as grand unified theories or Planck scale effects.
At this stage all couplings ${\bf Y}^{(i)}_{q,l}$, $\mu_{ij}$,
$\mu_{\Delta}$, $\mu_{\Delta^c}$, ${\bf f}$, ${\bf f}_c$ are 
complex with $\mu_{ij}$, ${\bf f}$ and ${\bf f}_c$ being symmetric matrices. 

The part of the supersymmetric action that arises from this
is given by
\begin{eqnarray}
{\cal S}_W = \int d^4 x \int d^2 \theta \, W + 
\int d^4 x \int d^2 \bar{\theta} \, W^\dagger \, .
\end{eqnarray}

The terms that break supersymmetry softly to make the theory 
realistic can be written as

\begin{eqnarray}
{\cal L}_{\rm soft} & = & \int d^4 \theta \sum_i m^2_i \phi_i^\dagger \phi_i
                      + \int d^2 \theta \, \theta^2 \sum_i A_i W_i 
     + \int d^2 \bar{\theta} \, {\bar{\theta}}^2 \sum_i A_i^* W_i^\dagger
                        \nonumber\\
     & + & \int d^2 \theta \, \theta^2 \sum_p m_{\lambda_p} 
                 {\tilde{W}}_p {\tilde{W}}_p +
           \int d^2 \bar{\theta} \, {\bar{\theta}}^2 \sum_p m_{\lambda_p}^* 
                 {{\tilde{W}}_p}^* {{\tilde{W}}_p}^* \, . 
\label{eq:soft}
\end{eqnarray}
In Eq. \ref{eq:soft},  ${\tilde{W}}_p$ denotes the gauge-covariant
chiral superfield that contains the $F_{\mu\nu}$-type terms with
the subscript going over the gauge groups of the theory including
SU$(3)_c$. $W_i$ denotes the various terms in the superpotential, 
with all superfields replaced by their scalar components and
with coupling matrices which are not identical to those in $W$.
Eq. \ref{eq:soft} gives the most general set of soft breaking terms
for this model.
          
It is clear from the above equations that this model has no baryon
or lepton number violating terms and if the ground state of the theory
does not break $B$ or $L$, then the LSP in this
model is stable and becomes the dark matter
particle. We will address the issue of the ground state of the theory 
in a subsequent section since the requirement of R-parity conservation
in the ground state requires that the model have in it nonrenormalizable
operators induced by higher scale physics such as Planck scale effects.

We noted earlier that left-right symmetry implies that the first
term in $\bar{\theta}$ is zero. Let us now see how supersymmetric 
left-right symmetry also requires the second term in this equation to 
vanish naturally. Under left-right 
transformation, the fields and the supersymmetric variable $\theta$
transform as follows:
\begin{eqnarray}
Q             & \leftrightarrow  &  {Q^c}^* \nonumber\\
L             & \leftrightarrow  &  {L^c}^* \nonumber\\
\Phi_i        & \leftrightarrow  &  {\Phi_i}^\dagger \nonumber\\
\Delta        & \leftrightarrow  &  {\Delta^c}^\dagger \nonumber\\
\bar{\Delta}  & \leftrightarrow  &  {\bar{\Delta}}^{c\dagger} \nonumber\\
\theta        & \leftrightarrow  &  \bar{\theta} \nonumber\\
{\tilde{W}}_{SU(2)_L} & \leftrightarrow  & {\tilde{W}}^*_{SU(2)_R} \nonumber\\  
{\tilde{W}}_{B-L,SU(3)_C} & 
         \leftrightarrow  & {\tilde{W}}^*_{B-L,SU(3)_C}
 \label{eq:lrdef}
\end{eqnarray}

Note that this corresponds to the usual definition 
$Q_L \leftrightarrow Q_R$, etc. 
With this definition of L-R symmetry, it is easy to check that

\begin{eqnarray}
{\bf Y}^{(i)}_{q,l} & = & {{\bf Y}^{(i)}_{q,l}}^\dagger \nonumber\\
\mu_{ij} & = & \mu_{ij}^* \nonumber\\
\mu_\Delta & = & \mu_{\Delta^c}^* \nonumber\\
{\bf f} & = & {\bf f}_c^* \nonumber\\
m_{\lambda_{SU(2)_L}} & = & m_{\lambda_{SU(2)_R}}^* \nonumber\\
m_{\lambda_{B-L,SU(3)_C}} & 
          = & m_{\lambda_{B-L,SU(3)_C}}^*\nonumber\\
A_i & = & A^\dagger_i,
\label{eq:rels}
\end{eqnarray}
       The first point to note is that the gluino mass is automatically
real in this model; as a result, the last term in the equation for
$\bar{\theta}$ above is naturally zero. We now therefore have to
investigate only the quark mass matrices in order to guarantee that
$\bar{\theta}$ vanishes at the tree level. For this purpose, we note that
the Yukawa matrices are hermitean and
the mass terms involving Higgs bidoublets in the superpotential are real.
If we can show that the vacuum expectation values of the bi-doublets are 
real, then the tree level value of $\bar{\theta}$ will be naturally zero.
For this purpose note that
the second observation above implies that the Higgs potential involving
the bidoublet and the triplet Higgs fields is completely CP conserving
(i.e. it has no complex parameters in it). This is encouraging since
a completely CP-conserving potential always has a domain of parameters
where the ground state is CP conserving. The story however is not so simple
since the soft breaking terms involving the sneutrinos (e.g. $\tilde{\nu}
 \phi \tilde{\nu^c}$) have a complex coupling and if the ground state
has a nonzero sneutrino vev, the ground state will break CP. A careful
study of the ground state is therefore essential for this purpose.

\section{Reality of the bidoublet vev's, tree level $\bar{\theta}$
and limit on the $W_R$ scale.}

To see whether the bidoublet vevs are real, one has to investigate the
ground state of the theory and in particular we have to see whether
the sneutrinos have vev. This was investigated by R. Kuchimanchi and this author
in Ref.\cite{kuch93}. It was noted there that the ground state of the
SUSYLR model does violate common intuition i.e. if only renormalizable
terms in the superpotential are kept in the minimal SUSYLR model, the
ground state does not break any of the gauge symmetries nor parity symmetry. 
This is easily seen in the supersymmetric limit since in this case
the mass-squares of the $\Delta$ Higgses are positive as are those for
the $\Phi$ fields and since the D-term contributions to the potential
are always positive, the above result follows. Once supersymmetry
breaks, in principle some of the above mass-squares can be negative
and the argument is not so simple. It however turns out that if a 
particular Higgs mass-square is negative, there are particular directions
in the Higgs space along which the D-term vanishes and in those
directions the negative nass-square will make the potential unbounded
from below. This means that no mass square should be allowed to be 
negative. This of course means that the symmetry cannot break. The
details of this argument are given in ref.\cite{kuch93}.

This result can 
however be avoided by adding a parity odd
singlet $\sigma$ to the model in which case one can write a superpotential
of the form 
\begin{eqnarray}
W(\sigma, \Delta ...)= \mu (\Delta\bar{\Delta}+\Delta^c\bar{\Delta^c})
+\lambda\sigma(\Delta\bar{\Delta}-\Delta^c\bar{\Delta}^c) + m_{\sigma}\sigma^2
\end{eqnarray}
which after supersymmetry breaking soft terms gives rise to parity violating
and $SU(2)_R$ violating minimum. The problem now is that\cite{kuch93}
is that the ground state of this slightly extended model
is also quite unusual in that it breaks electric charge unless R-parity
is spontaneously broken by the $<\tilde{\nu^c}>\neq 0$ . But this is precisely
what we wanted to avoid in order to get real bidoublet vev's and solve the
strong CP problem. How is this
to be achieved ? It was pointed out in \cite{moha96} that if one adds
nonrenormalizable Planck scale induced terms to the superpotential, one can
indeed have a parity violating vacuum that conserves R-parity. Since this
is such an important point, we review this proof in an appendix.
The conclusion of this part then is that if one includes non-renormalizable
Planck scale induced terms to the minimal SUSYLR model, the ground state
corresponds to having zero sneutrino vev's and therefore all couplings
that zeroth order in $\frac{M_{wk}}{M_{Pl}}$ in the Higgs potential are
real leading to real bidoublet vevs to this order. Note however that
one must include all allowed nonrenormalizable terms to the model and one such
term has the form $\phi\phi\Delta^c\bar{\Delta}^c/M_{Pl}$ which can
come with a complex coupling. This will then induce a tree level phase
in the bidoublet vevs of order $v^3_R/M^2_SM_{Pl}$\cite{goran97}. For its
effect on $\bar{\theta}$ to be less than $10^{-10}$, one must have
$v_R\leq 10^{5}$ GeV or so. 

To summarize this sub-section, the requirement of tree level value of
theta to be naturally small implies that there must be non-renormalizable
terms in the superpotential and the scale of the right handed $W_R$ must be
in the few TeV range.

\section{One-loop contributions to $\bar{\theta}$ and the elctric dipole
moment of the neutron}

Having shown that the tree level contributions to the $\bar{\theta}$ are
small, we now proceed to discuss the magnitude of the one loop contributions
to the quark and gluino masses.
There are two classes of diagrams that have to be considered for the quark
masses: one where the gluino intermediate
state is present and another where the Wino states contributes. Let us
first discuss the gluino effect given in the Fig.2. 

\subsection{Gluino contribution}

We work for simplicity in the case in which thevevs of $\Phi_a$
are in the form $Diag<\Phi_1>=(\kappa_u,0)$ and $Diag <\Phi_2>=(0,\kappa_d)$.
Note that due to left
right symmetry the gluino mass is real. Any complexity of the determinant of
the quark mass matrices must therefore come from the supersymmetry breaking
contributions to the squark masses. The general form of the squark mass
matrix for the down sector can be written as\cite{posp96}:
\begin{eqnarray}
M^2_{\tilde{d}\tilde{d}^c}=\left(\begin{array}{cc}
m^2_L + M^2_d +c_u h^2_u  &  X \\
X^{\dagger} & m^2_R + M^2_d+ c'_u h^2_u \\
\end{array} \right)
\end{eqnarray}
where $X= (A-\mu tan\beta)(M_d + a_u M_dh^2_u +h^2_uM_d a'_u)$.We have
ignored small contributions proportional to the down quark couplings.
 Note that
in the limit of exact left-right symmetry, $c_u=c'_u$, $a_u=a'_u$.    
However, after parity breaking effects are included, we have
$c_u-c'_u \simeq \frac{g^2}{16\pi^2} ln \frac{m^2_{W_L}}{m^2_{W_R}}$, etc.
In order to calculate the one loop contribution to the quark edm, we
focus on the down quark and treat the $c$ terms, the $M^2_d$
 and the $X$-terms as 
perturbations. To zeroth order in this perturbation,
 then we have a diagram which is given
in figure 2 but without the solid circles which represent the insertion
of $c_u$ and $c'_u$ terms. Also let us work in a basis in which the up
quark mass matrix is diagonal.It is then easy to see that, the loop is 
proportional to $a_u M_dh^2_u+a'_uh^2_uM_d$. Since $M_d$ is hermitean and
$M^2_u$ is diagonal, the $11$ element of this matrix is real and therefore
it makes no contribution to the edm of the down quark. In order to see
at what level we will get a contribution to the quark edm we keep inserting the
various perturbation terms proportional to $c_u$ and $c'_u$ until we
get a complex $11$ entry for the down quark mass matrix. It turns out
that the dominant contribution to the mass matrix is given by\cite{posp96}:
\begin{eqnarray}
M^{1-loop}_d\simeq Im(V^*_{td}V_{tb}V^*_{cb}V_{cd})\frac{\alpha_s}{4\pi}
\frac{3\alpha_W}{2\pi}h^2_ch^4_tm_bZ_{susy} ln\frac{M^2_{W_R}}{M^2_{W_L}}
\end{eqnarray}
 where $Z_{susy}=\frac{m_{\tilde{G}}(A-\mu tan\beta)}{M^2_{susy}}$.
The fact that this CP-violating contribution would be proportional to
the product of the CKM mixing angles was noted in \cite{moha96} from
simple analytic arguments. In ref.\cite{posp96}, an enhancement of the
result by a factor $m_b/m_d$ was noted. Thus the contribution of this
to the $\bar{\theta}$ is:
\begin{eqnarray}
\bar{\theta} \simeq Im(V^*_{td}V_{tb}V^*_{cb}V_{cd})\frac{\alpha_s}{4\pi}
\frac{3\alpha_W}{2\pi}h^2_c\frac{m_b}{m_d} Z_{susy}
ln\frac{M^2_{W_R}}{M^2_{W_L}}
\end{eqnarray}
where we have set $h_t\simeq 1$ as is suggested by the recent top quark mass
measurements. Putting in all the numbers, one finds that $\bar{\theta}
\simeq 10^{-9}$. Barring unforeseen cancellation of parameters therefore,
this will lead to a value for the electric dipole moment of the neutron
of order $10^{-25}$ to $10^{-26}$ ecm for $W_R$ masses in the TeV range.
This number is quite close to the present upper limit. At present there
are plans to improve the search for the edm of neutron by several orders
of magnitude\cite{lam96} by using ultra cold neutron in superfluid He$^4$.
These experiments will therefore provide a decisive test of this class
of supersymmetric left-right models.

\subsection{One loop wino contributions and C invariance}

Next let us consider the wino contributions to the quark masses. A typical
graph is shown in Fig.3. We see from this that it involves the majorana
mass for the $SU(2)_L$ Wino . There is a similar graph involving the $SU(2)_R$
wino. If parity was an exact symmetry and the full mass of the left and the
right winos were the same, these two two graphs would have added up to give
a real quark mass at the one loop. However since parity breaking leads to
a non-cancellation between these two diagrams, we will get a contribution
to quark masses of order $\simeq \frac{\alpha_w}{4\pi}M_{\lambda_w}$. In
theories where the wino and zino masses vanish, this contribution is 
suppressed. However in general, this will be present. Again it may turn
out that in specific theories, this mass may be real (as in some gauge
mediated supersymmetry breaking models) in which case again, this keeps
the tree level hermiticity of the mass matrix and the contribution to 
$\theta$ vanishes. In general however, the dynamical content of the theory
may be such that neither happens. In this case, it is necessary to
impose in addition a charge conjugation (C) symmetry on the Lagrangian
which makes these contributions vanish\cite{goran97}. It actually turns
out that the C-symmetry must be softly broken to keep the CKM phase
in weak current from vanishing.  

\subsection{One loop phase in the gluino masses}

 A rough order of magnitude of the CP violating
phase in the gluino mass can be estimated as follows: since the 
${\bf A}_{u,d}$ are hermitean and proportional to the Yukawa couplings 
${\bf h}_{u,d}$
at some scale above the $M_R$ scale, let us go to a basis where 
${\bf h}_d$ and ${\bf A}_d$ are diagonalized. Then we find that, at the scale
of proportionality, if any one of 
the off-diagonal elements of ${\bf h}_u$ and (hence ${\bf A}_u$) are set to 
zero, the theory becomes
completely CP conserving and cannot generate a CP violating phase 
at any scale below $M_R$. It is then clear that the one loop graph
that generates a phase in the gluino mass can lead to the gluino phase
$\delta_{\tilde{g}}$ which is at most
\begin{eqnarray}
\delta_{\tilde{g}}\simeq {{V_{ub}V_{bc}V_{cd}V_{du}\alpha_s
\alpha_w}\over{8 \pi^2}}Z_{susy}\frac{m^2_b}{M^2_{\tilde{G}}}
ln{{M^2_R}\over{M^2_Z}}
\end{eqnarray}
leading to $\delta_{\tilde{g}}\leq 10^{-11}$ which is negligible.
 
Thus the conclusion of this section is that the finite values of 
$\bar{\theta}$ induced at the one loop level are within the present
experimental limits so that SUSYLR model indeed provides a satisfactory
solution to the strong CP problem. The SUSYCP problem is then of course
automatically solved.

\vskip 0.3cm
                       
\section{Phenomenological Implications: light doubly charged Higgs bosons}

The low $W_R$ left-right models are rich in phenomenology as is well known.
Apart from the obvious searches for the new particles of the model such as
the $W_R$, $Z_{LR}$ and the heavy right-handed neutrino, there are a large
number of new processes induced by the non-standard model particles present.
A very typical example of such process is the neutrinoless double beta
decay induced in these models by the exchange of right-handed neutrinos
\cite{moh86}, doubly charged Higgs bosons\cite{mohverg82} etc. In fact,
the recent stringent lower limits on the lifetime for neutrinoless
double beta decay in enriched $^{76}$Ge by the Heidelberg-Moscow 
collaboration\cite{klap} implies that $M_{W_R}\geq 1.1$ TeV and
$M_{\nu_R}\geq 1.1$ TeV when combined with theoretical limits from
vacuum stability. 
There are also comparable stringent limits on the $W_R$ mass of the order
of a TeV from the $K^0-\bar{K}^0$ mass difference\cite{beall}. The bidoublet
Higgs induced flavor changing neutral currents also imply that the all but
the lightest standard model Higgs boson must be in the 5 to 10 TeV 
range\cite{po96}. There are also strong limits on the charged Higgs 
masses\cite{gautam2} and the $W_L-W_R$ mixing angle\cite{babu} 
from the $b\rightarrow s\gamma$ measurements.
 Finally there are recent collider limits on the mass
of the $W_R$ and $Z_{LR}$\cite{cvetic} 
from the D0 and CDF experiments\cite{abe}. The
detailed discussion of these topics is beyond 
the scope of this talk.

Another very striking prediction of these models is the existence of
light ( 100 GeV to TeV range) doubly charged Higgs bosons. These lead to
many interesting phenomenological implications such as muonium-antimuonium
transition\cite{halp} on which there now exist the most stringent upper
limit of $\leq 3\times G_F 10^{-3}$ from the PSI\cite{jung}.
This implies that the combination of couplings and masses of the doubly
charged bosons is restricted as follows: $\frac{f_{ee}f_{\mu\mu}}{M^2_{\Delta}}
\leq 2\times 10^{-7} GeV^{-2}$. 
This particle also contributes to the g-2 of the muon. The present measurements of g-2 imply that the additional contribution from
the doubly charged Higgs boson must be bounded above by $8.5\times 10^{-9}$
implying that 
$\frac{f^2_{\mu\mu}}{M^2_{\Delta}}\leq 1.4\times 10^{-4} GeV^{-2}$
The present BNLE821 experiment aims to push the accuracy to the level of
$4\time 10^{-10}$ level making the above constraint more stringent by a
factor of 21.
 They also have
many implications for collider phenomenology.

There have also been study of the new supersymmetric particles of these
models, specially in regard to their manifestation in colliders\cite{kalman}.
The literature in the subject is vast and we do not discuss it any further.
Instead, I focus on the recent work on trying to explain the HERA anomaly
in terms of the leptoquarks that arise naturally in the context of the
SUSYLR models in the next section.

\section{Leptoquarks and HERA anomaly in the SUSYLR model}

If the high $Q^2$ anomaly observed recently in the $e^+p$ scattering by the
H1\cite{H1} and the ZEUS\cite{ZEUS} collaborations is confirmed by future data,
it will be an extremely interesting signal of new physics beyond the standard
model. A very plausible and widely discussed interpretation of this anomaly
appears to be in terms of new scalar particles capable of coupling to $e^+u$ or
$e^+d$ of scalar leptoquarks\cite{lq} with mass around 
200 GeV. Alternative interpretations based on contact interactions
\cite{contact} or a second $Z'$\cite{godfrey} have been proposed; but
attempts to construct models that lead to the desired properties
seem to run into theoretical problems.
 
The leptoquark
must be a spin zero color triplet particle with electric charge 
$5/3$ or $2/3$. The latter electric charge assignment makes it possible
to give a plausible interpretation of the leptoquark as being the 
superpartner of the up-like quark\cite{squark} of
the minimal supersymmetric standard model (MSSM) provided one includes
the R-parity violating couplings of the type $\lambda' QLd^c$ to the MSSM
 There are however very stringent
upper limits on several R-parity violating couplings: for instance, if
in addition to the $\lambda'$ term described above, one adds the 
allowed $\lambda''u^cd^cd^c$ terms to the superpotential, then it leads to
catastrophic proton decay unless $\lambda'\lambda''\leq 10^{-24}$
\cite{gautam}. There are also stringent limits on $\lambda'_{111}\leq 10^{-4}$
\cite{mhkk} from neutrinoless double beta decay, which forces the
lepto-quark to be $\tilde{c}$ or $\tilde{t}$ rather than the obvious choice
$\tilde{u}$. Furthermore, within such a framework, the lightest
supersymmetric particle (LSP) is no more stable and therefore, there is
no cold dark matter (CDM) candidate in such theories.
Since supersymmetric left-right lead to automatic R-parity conservation 
we have recently used them as a typical framework for studying the consequences
of leptoquarks without giving up R-parity conservation\cite{satya}. 
We augment the SUSYLR model by including the leptoquark fields. 
Demanding that the lepto-quarks couple to $e^+d$ or $e^+u$ 
leads to the conclusion that
they must belong to the multiplet $(2,2, 4/3,3^*)$ (denoted $\Sigma_{QL^c}$).
Anomaly cancellation requires that there be a conjugate state
$(2,2,-4/3, 3^*)$ (denoted by $\bar{\Sigma}_{Q^cL}$). Each of these multiplets
have four scalar fields and four fermion fields which will be denoted
in what follows by the obvious subscript corresponding to their couplings.
We denote the four scalar leptoquarks as
$\bar{\Sigma}_{ue^c}$, $\bar{\Sigma}_{u\nu^c}$, $\bar{\Sigma}_{de^c}$
and $\bar{\Sigma}_{d\nu^c}$ and their fermionic partners (to be denoted
by a tilde on the corresponding scalar field).
Writing down the most general superpotential, one can easily convince oneself
that the resulting theory maintains the property of automatic R-parity
conservation. 

Before discussing the application of this model to explain the HERA
anomaly, let us first discuss the mass spectrum of the model. The
superpotential for this model will have a direct mass term of the form
$M_0\Sigma \bar{\Sigma}$ which will imply that the fermionic fields
in the leptoquark multiplet will have a masses $M_0$ prior to symmetry
breaking. There may be other contributions to these masses from radiative
corrections which will split the degeneracy implied by the above mass term.

As far as the scalar leptoquark states are concerned, their masses will
receive several contributions: 
first a direct common contribution from the $M_0$
term given above. After symmetry breaking, the D-terms of the various
gauge groups will contribute. There is also soft SUSY breaking contribution
along with the radiative correction. Assuming that the
$SU(2)_R$ symmetry is broken by the vev's $<\Delta^c>=v_1$,
$<\bar{\Delta}^c>=v_2$ and the
$SU(2)_L$ symmetry is broken by the two $\phi$ vev's as $diag <\phi_u>=
(0,vsin\beta/\sqrt 2 )$ and $diag<\phi_d>=(vcos\beta/\sqrt 2,0)$, 
we can write the masses for the various scalar leptoquarks as follows:
\begin{eqnarray}
M^2_{\Sigma_a}&=& M^2_0 + (I^a_Rg^2_{2R}-\frac{B-L}{8}g^2_{B-L})(v^2_1-v^2_2)
\\\nonumber&+&\frac{g^2_{2L}v^2}{4} cos 2\beta+\Delta_m^2+{\rm Radiative\,
correction}
\end{eqnarray}
The values of the $I^a_R$ and $B-L$ are given in Table I:

\begin{table}
\begin{tabular}{|c||c||c||c|} \hline
states & $I^a_R$ & $B-L$ & $I_{3L}$ \\ \hline
$u^ce$ & $\frac{1}{2}$ & $\frac{4}{3}$ & $\frac{1}{2}$ \\\hline
$u^c\nu$ & $\frac{1}{2}$& $\frac{4}{3}$& $-\frac{1}{2}$ \\\hline
$ue^c$ & $-\frac{1}{2}$ &$-\frac{4}{3}$ & $-\frac{1}{2}$ \\\hline
$de^c$ & $-\frac{1}{2} $ & $-\frac{4}{3}$ & $\frac{1}{2}$ \\\hline
$d^c\nu$ & $-\frac{1}{2}$ & $\frac{4}{3}$ & $-\frac{1}{2}$\\\hline
$d^ce$ & $-\frac{1}{2}$ & $\frac{4}{3}$ & $-\frac{1}{2}$\\ \hline
\end{tabular}
\caption{The $I_{3R}$ and $B_L$ quantum numbers of the various leptoquark
states }
\end{table}
 The radiative correction can  be positive or negative.
From table I and Eq.(1), we see easily that if $v_1 < v_2$, then the lightest
leptoquark state is $\Sigma_{u^ce}$ since $g^2_{2R}> 2g^2_{B-L}$ in the
left-right models. Furthermore, interestingly enough for
this choice of vev's, assuming the combined  $\Delta_m^2+{\rm Radiative\,
correction}$ to be smaller
than the the D-term, the leptoquarkino states are heavier than the lightest
leptoquark state.  We assume their masses to be in the range of 300 to 400 GeV. 
In a subsequent section we will obtain a lower bound on
the leptoquarkino mass from the present collider data.

Turning to the couplings of the leptoquarks $\Sigma $ and $\bar{\Sigma}$
to quarks and leptons, it is given by the superpotential:
\begin{eqnarray}
W_{lq}= \lambda_{ij} (\Sigma Q^c_iL_j +\bar{\Sigma} Q_i L^c_j)
\end{eqnarray}
(where $i,j$ are generation indices). Let us assume for simplicity that 
$\lambda_{ij}$ are diagonal in the mass eigenstate basis for the quarks and
leptons. Explanation of the HERA high $Q^2$
anomaly seems to require $\lambda_{11}\simeq 0.05$ which we will
assume from now on and mass of the scalar field in $\Sigma$ (assumed to
be $\Sigma_{u^ce}$ from the above mass arguments) around 200 GeV. 
As far as the other couplings go, they will be strongly constrained
by the present experimental upper limits on the low energy processes 
such as $\mu\to e\gamma$, 
$\tau\to e\gamma$ etc. which will arise at the one loop level from the exchange
of $\Sigma$ and $\bar{\Sigma}$. There are also tree level diagrams
which can lead to rare processes such as $K\to \pi e^-\mu^+$. These 
processes imply an upper limit of $\lambda_{22}\leq 2\times 10^{-3}$.
The most stringent upper limit on $\lambda_{22}$ comes from the upper
limit on the process $K^0_L\to \mu^+e^-$ and yields $\lambda_{22}\leq 2\times
10^{-4.5}$. Turning to $\lambda_{33}$,
the most stringent limits arise from the present upper limit
on the branching ratio for the process $\tau\to e\gamma$  which
the 1996 Particle data tables give as $\leq 1.1\times 10^{-4}$.
This implies a weak upper limit on $\lambda_{33}\leq 0.2$. Thus the third
generation leptoquark coupling $\lambda_{33}$ could in principle be comparable
to $\lambda_{11}$. If the efficiency for the detection of $\tau$ leptons
at HERA were comparable to the detection efficiency for electrons, that
could also severly limit $\lambda_{33}$. 
Finally we also note that such a value for $\lambda_{11}$
is also consistent with the present data on the parity violation in atomic
physics. Thus it appears that our leptoquark couplings are consistent with
all known low energy data.

An important consequence of the above coupling is that the branching ratio
for the leptoquark coupling to the electron mode is only 50\%. As a result
the recent CDF/D0 bound for the leptoquark mass is around 195 GeV\cite{D0}
which is lower than mass required for explaining the HERA anomaly.
                   
The existence and properties of the leptoquarkino provides a new and 
unique signature for our model as compared to all other proposals to 
explain the HERA anomaly. To see this note that in hadron colliders,
we can produce pairs of leptoquarkinos at the same rate as the $t\bar{t}$
pair due to identical color content. Moreover, due to R-parity conservation
the leptoquarkino decay leads to a missing energy signal as follows:
$\tilde{\Sigma}\to e^+\tilde{u}$ with $\tilde{u}\to u + \chi^0$ or
$\tilde{\Sigma}\to u+\tilde{e^+}$ with $\tilde{e^+}\to e^+ + \chi^0$. In both
the cases we have $e^+,u$ plus missing energy in the final state. If the
$\lambda_{33}$ coupling is comparable to $\lambda_{11}$ as is allowed,
then the branching ratio for leptoquarkino decay to electrons 
will be 50\%. This signal is similar
to the top signal at the Tevatron with one crucial difference. 
The dilepton branching ratio from the leptoquark pairs is 100$\%$ compared to
only 10$\%$ from the top pairs. There will be no $\mu^{+} \mu^{-}$ events. The
branching ratio to $e^{+} e^{-}$ will be 25$\%$ compared to only 1$\%$ from the
top. The present observations should therefore lead to lower limits on the mass
of the lightest leptoquarkino pairs. Thus an excess of dilepton pairs over that
expected from the top productions, or an excess in $e^{+} e^{-}$ channel over
$\mu^{+} \mu^{-}$ will be a clear signal of leptoquark productions at Tevatron.
In addition, leptoquarks will also give rise to harder 
leptons and larger missing
energy events.
 One could also look for the signals
of leptoquarkino production in
$e^+\gamma$ colliders. The leptoquark or leptoquarkino can be singly produced in
such a  collider. The scalar leptoquark will be produced in the association of 
an antiquark. As discussed before in the usual SUSY theories, the
leptoquark further decays  into a quark and a positron with 
the final state consisting of positron +jets. The leptoquarkino however will be
 produced along with a squark.  This leptoquarkino will then decay into a 
lepton and a squark. The final state has electron +jets+missingenergy. 
The missing energy part will then disentangle the leptoquarkino signal. In the
gauge mediated SUSY breaking scenario, the final state has either a hard
photon or $\tau^+\tau^-$ along with electron +jets+missingenergy.
              
Another important implication of our model is that at the renormalizable
level, there are
any operator which can give a non-negligible charged current signal at HERA
without conflicting with $\pi^+\rightarrow e^+\nu_e$ decay constraint.
Thus if a significant amount of charged current like events are observed
at HERA this model will have to be further extended.

\vskip 0.6cm

\section{Grandunification with low mass $W_R$}

There has been long standing interest in the question of whether low scale
supersymmetric left-right models lead to unification of couplings. Several
examples of low scale scale left-right models that grand unify already
exist in the literature\cite{desh}. But all of them lead to arbitrary
R-parity violation and are therefore outside the philosophical framework
of this paper. It is interesting that it is possible to find a low mass
$W_R$ SUSYLR model which leads to grand unification. The basic strategy
was to observe\cite{brahm} that if in given model, the
combination of the one loop beta function 
coefficients:
\begin{eqnarray}
 \Delta b\equiv 5 b_1 - 12 b_{2L} +7 b_3
\end{eqnarray}
 vanishes, 
then the new physics scale in this model can be low consistent with
the present precision measurements of the gauge couplings. It turns out that
if the parity symmetry (but not the $SU(2)_R$) is brioken at the GUT scale,
then one can choose the following spectrum of particles above the $M_R$ scale
to have the $\Delta b=0$ and thereby a $W_R$ mass in the TeV range: the
multiplets are : two bidoublets $\Phi (2,2,0)$; the right handed triplets
that lead to the see-saw mechanism: $\Delta^c (1,3 +2)+\bar{\Delta}^c
(1,3,-2)$, a color octet field neutral under the electroweak gauge group and
one $B-L$ neutral $SU(2)_L$ triplet. The one loop evolution equations for
the gauge couplings in this case is displayed in Fig.4. The unification
scale is safely in the range of $10^{16}$ GeV or so. The important point
for our discussion is that this model respects automatic R-parity conservation
and also solves the SUSY CP problem as discussed in the text. Whether this
model also solves the strong CP problem is not clear yet and is presently
under investigation. The unification group is $SO(10)$ or any other group
of which $SO(10)$ is a subgroup.

\section{Conclusion}

In this talk we have shown that the class of supersymmetric left-right models
that implement the see-saw mechanism for neutrino masses have two very
desirable feature: (i) automatic R-parity conservation and (ii) solution to
the susy CP problem due to constraints of parity symmetry. These results are
independent of the scale of the left-right symmetry breaking. On the other hand
if the $W_R$ scale is in the TeV range, one also has a rather elegant solution
to the strong CP problem without the need for an axion. A crucial prediction
of this class of models that solve the strong CP problem is that the electric
dipole moment of the neutron is of order $10^{-25}$ to $10^{-26}$ ecm which
is measurable in the next round of proposed experiments. Finally we discuss
how the SUSYLR model can incorporate the leptoquarks without the need for
R-parity violation in order to explain the HERA high $Q^2$ anomaly.

\vskip 0.6cm

\noindent{\bf APPENDIX:  Avoiding Sneutrino VEVs}

\vskip 0.6cm

In this Appendix we will show that if in the minimal SUSY LR model one 
includes non-renormalizable Planck scale induced terms, the ground state
of the theory can be $Q^{em}$ conserving even for 
$<{\tilde{\nu}}^c> = 0$. For this purpose, let us briefly recall the argument
of Ref. \cite{kuch93}. The part of the potential containing 
${\tilde{L}}^c$, $\Delta^c$ and ${\bar{\Delta}}^c$ fields only has 
the form (see Appendix B or \cite{kuch93} )

\begin{eqnarray}
V = V_0 + V_D \, ,
\end{eqnarray}

where

\begin{eqnarray}
V_0  = 
{\rm Tr} | i {\bf f}^\dagger L^c {L^c}^T \tau_2 + 
\mu_\Delta^* \bar{\Delta}^c|^2 \nonumber\\
 + 
\mu^2_1 {\rm Tr} ( \Delta^c \Delta^{c\dagger}) \, +
\mu^2_2 {\rm Tr} ( {\bar{\Delta}}^c {\bar{\Delta}}^{c\dagger} ) \nonumber\\
 + 
\mu^2_3 {\rm Tr} (\Delta^c \bar{\Delta}^c ) +
\mu_4 {\tilde{L}}^{cT} \tau_2 \Delta^c L^c \, ,  
\end{eqnarray} 

and

\begin{eqnarray}
V_D  =  
{g^2 \over 8} \sum_{m} | {\tilde{L}}^{c\dagger} \tau_m {\tilde{L}}^c +
{\rm Tr} ( 2 {\Delta}^{c\dagger} \tau_m {\Delta}^c + 
     2 {\bar{\Delta}}^{c\dagger} \tau_m {\bar{\Delta}}^c  |^2 \nonumber\\
   + 
{g'^2 \over 8} | {\tilde{L}}^{c\dagger} {\tilde{L}}^c -
2 \, {\rm Tr} ( {\Delta}^{c\dagger} {\Delta}^c -\bar{\Delta}^{c\dagger}
\bar{\Delta^c} )|^2
\end{eqnarray}

Note that if $<{\tilde{\nu}}^c> = 0$ then the vacuum state for which
$\Delta^c = { 1 \over \sqrt{2} } v \tau_1 $ and
${\bar{\Delta}}^c = { 1 \over \sqrt{2} } v' \tau_1 $ is 
lower than the vacuum state
$
\Delta^c = v
\left (
\begin{array}{cc}
0 &  0 \\ 
1 &  0 \\
\end{array} \right ) $ and
$
{\bar{\Delta}}^c = v'
\left (
\begin{array}{cc}
0 &  1 \\ 
0 &  0 \\
\end{array} \right ) \, .
$
However, the former is electric charge violating. The only way to 
have the global minimum conserve electric charge is to have
$<{\tilde{\nu}}^c> \neq 0$. On the other hand, if we have 
non-renormalizable terms included in the theory, the situation changes:
for instance, let us include non-renormalizable gauge invariant terms of the
form (inclusion of other non-renormalizable terms simply enlarges the
parameter space where our conclusion holds):

\begin{eqnarray}
W_{NR} = {\lambda \over M} 
[ {\rm Tr} ( \Delta^c \tau_m {\bar{\Delta}}^c) ]^2 \, .
\end{eqnarray}
This will change V to the form:

\begin{eqnarray}
V = V_0 + V_{NR} + V_D \, ,
\end{eqnarray}
where $V_0$ and $V_D$ are given before and $V_{NR}$ is given by
\begin{eqnarray}
V_{NR} = {{\lambda \mu} \over M} 
[ {\rm Tr} ( \Delta^c \tau_m {\bar{\Delta}}^c )]^2
+ {{4 \lambda \mu_\Delta} \over M} 
[ {\rm Tr} ( \Delta^c \tau_m {\bar{\Delta}}^c )]
[ {\rm Tr} ( \Delta^{c\dagger} \tau_m \Delta^c ) ]
+ \Delta^c \leftrightarrow {\bar{\Delta}}^c + {\rm etc.} 
\end{eqnarray}
For the charge violating minimum above, this term vanishes but the 
charge conserving minimum receives a nonzero contribution. 
Note that the sign of $\lambda$ is arbitrary and therefore, by 
appropriately choosing  sgn($\lambda$) we can make the
electric charge conserving vacuum lower than the $Q^{em}$-violating one.
In fact, one can argue that, since we expect
$v^2 - v'^2 \approx { {f^2 (M_{SUSY})^2} \over {16\pi^2} }$ in typical 
Polonyi type models, the charge conserving minimum occurs for
$ f < 4 \pi 
\left( { {4 \lambda \mu_\Delta} \over M } \right)^{1 \over 4}
{{v}\over{M_{SUSY}}}$.
For $\lambda \approx 1$, $\mu_\Delta \approx v \approx M_{SUSY} 
\approx 1 {\rm TeV}$ and $M= M_{Pl}$,
we get $f \leq 10^{-3}$ if $v-M_{SUSY}$. 
We have assumed that the right handed scale
is in the TeV range.
 The constraint on $f$ of course becomes weaker for 
larger values of $\mu_\Delta$.

\vskip 0.3cm

\newpage
\begin{center}
{\bf Acknowledgement}
\end{center}

I would like to thank Prof. Hans Klapdor-Kleingrothaus for kind 
hospitality at the Ringberg castle and for creating a warm and
pleasant atmosphere for scientific discussions.
This work has been supported by the National Science 
Foundation grant no. PHY-9421386. I would like to thank B. Brahmachari
for help with drawing Fig.4 and discussions on the supersymmetric
grand unification, B. Dutta, S. Nandi for discussions and collaboration
on the leptoquarks and A. Rasin and G. Senjanovi\'c for many
discussions and collaboration on the strong CP problem.

\noindent{\bf Figure Caption}

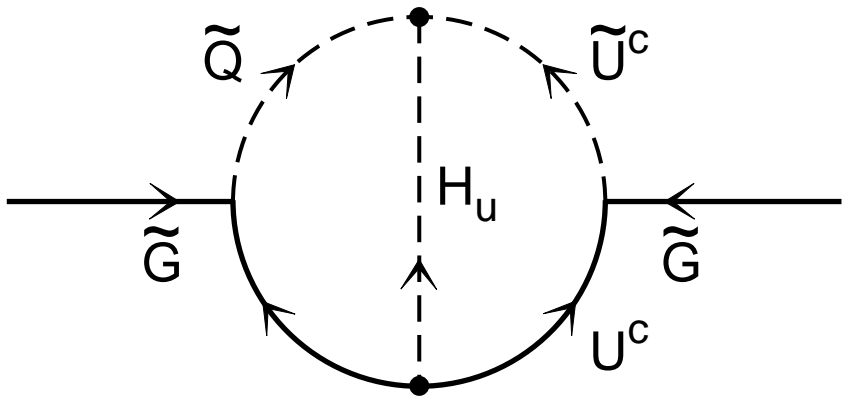
\noindent {\bf Figure 1:} A typical two loop graph that contributes to the
infinite phase of the gluino mass.

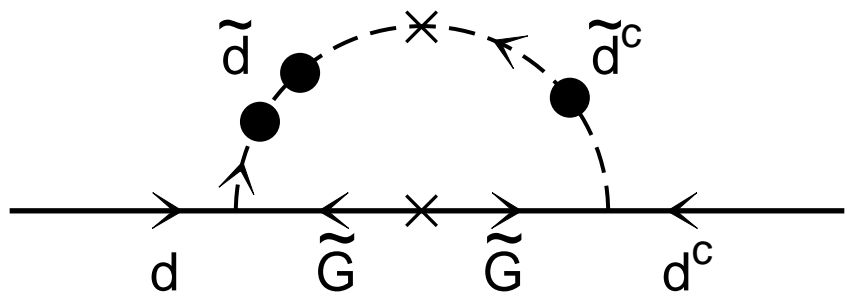
\noindent{\bf Figure 2:} One loop contribution to $\theta$ from the gluino
virtual state. The solid circles represent the squark mass insertions
proportional to $h^2_u$.

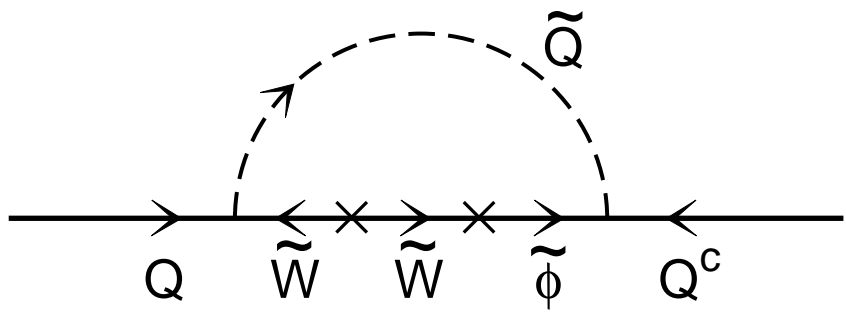
\noindent{\bf Figure 3:} One loop wino contribution to $\theta$ if the
wino masses are complex.

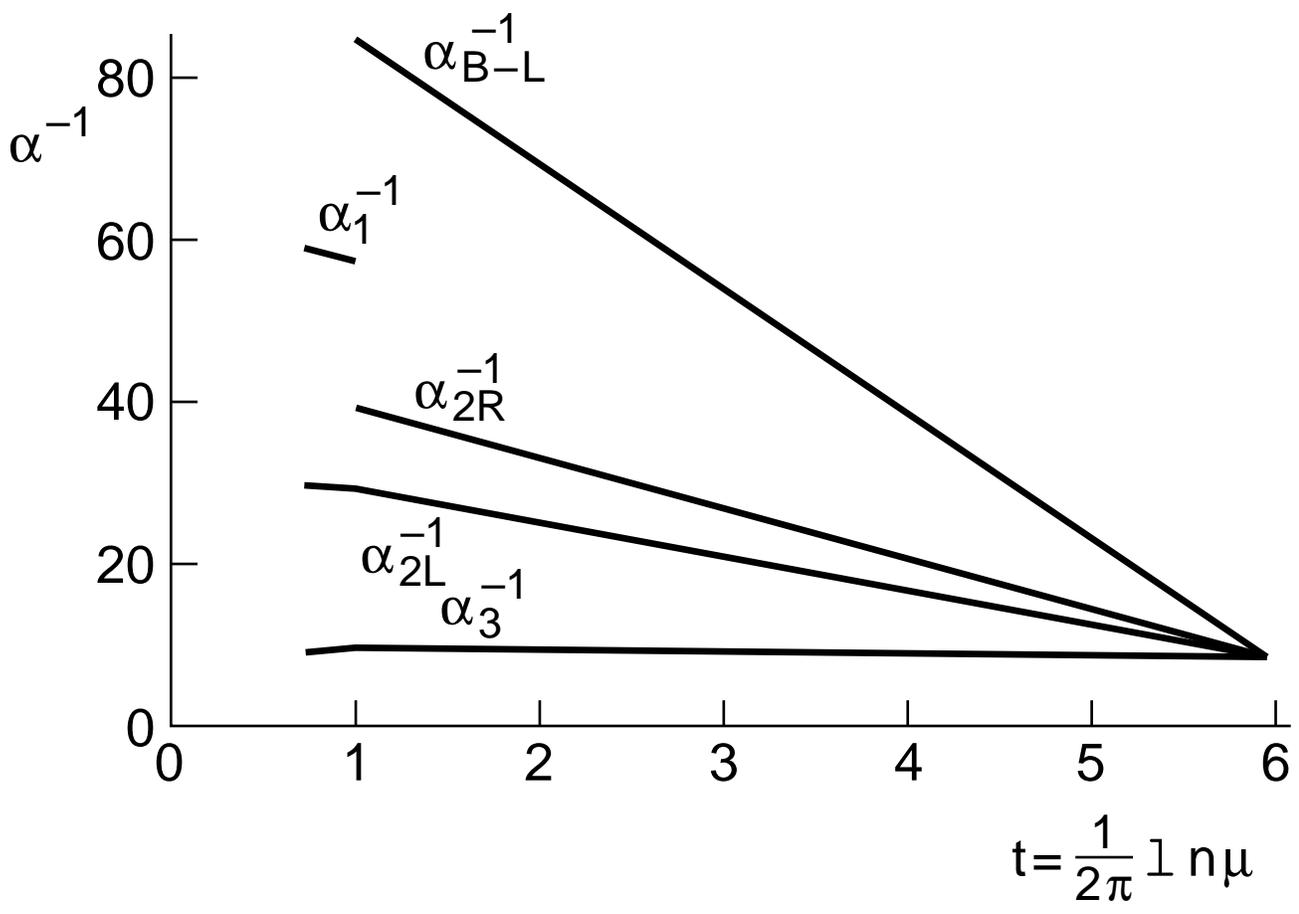
\noindent{\bf Figure 4:} Gauge coupling unification with the $M_{W_R}$ in
the TeV range.

\end{document}